# A Mobile Cloud Collaboration Fall Detection System Based on Ensemble Learning

Tong Wu, Yang Gu, Yiqiang Chen, Yunlong Xiao, and Jiwei Wang

Institute of Computing Technology, Chinese Academy of Sciences, Beijing, China

Email: {wutong17s, yqchen, guyang, xiaoyunlong, wangjiwei}@ict.ac.cn

*Abstract*—Falls are one of the important causes of accidental or unintentional injury death worldwide. Therefore, this paper presents a reliable fall detection algorithm and a mobile cloud collaboration system for fall detection. The algorithm is an ensemble learning method based on decision tree, named Fall-detection Ensemble Decision Tree (FEDT). The mobile cloud collaboration system can be divided into three stages: 1) mobile stage: use a light-weighted threshold method to filter out the activities of daily livings (ADLs), 2) collaboration stage: transmit data to cloud and meanwhile extract features in the cloud, 3) cloud stage: deploy the model trained by FEDT to give the final detection result with the extracted features. Experiments show that the performance of the proposed FEDT outperforms the others' over 1-3% both on sensitivity and specificity, and more importantly, the system can provide reliable fall detection in practical scenario.

## I. INTRODUCTION

According to the World Health Organization, falls are the second leading cause of accidental or unintentional injury death worldwide. Injures caused by falls are among elderly and are a major cause of permute death, disability, pain, and loss of independence. Each year, there are about 646,000 fatal falls occurred. Roughly 28-35% of people who are more than 65 years old fall and the percentage increases to 32-42% for those more than 70 years old [1].

As for these problems, many research works about fall detection have been carried out recently. These methods can be categorized into vision-based, ambient sensor based, and wearable sensor based [2][3]. The vision-based methods are popular in recent years due to the developments of computing ability and deep learning. They use cameras to capture images or videos of the people in a fixed scene, and through processing these images and videos they can detect falls [4]. The ambient sensor based methods collect information of the surroundings such as Wi-Fi [5], ultrasonic [6], radar [7], infrared ray [8]. Through analyzing signals from ambient sensors, they give detection results of falls. The wearable sensor based methods use the inertial sensors, such as accelerometer and gyroscope [9]. Combining with the machine learning algorithms, they can distinguish falls and ADLs. However, there are defects of the vision-based methods and ambient sensor based methods. Both in vision-based methods and ambient sensor based methods, they are less flexible for that the detection devices of them need to be placed in a fixed area, and the privacy problem is concerned as they are not target-specific and collect all information of people in the area. What's more, vision-based methods are costly, as storing and processing the images and videos needs large storage and high performance computer. However, the wearable sensor based methods can avoid these defects. The wearable sensor can be embedded into wearable devices, such as watches [10], eyeglasses [11], smart phone [12], shoes [13]. Additionally, they are target-specific for that they just record information of device owners. In this paper, we propose an algorithm with high sensitivity and specificity for fall detection, and implement a mobile cloud collaboration system with a wearable device and cloud server.

The main challenges of fall detection with wearable devices are low sensitivity, high false positive rate of fall detection in daily living activities and less robustness for different subjects and devices. Due to the limited computing resources of wearable device, the offline models in the wearable device are usually too simple to give high performance of sensitivity or less false positive rate of fall detection. Besides, as the models are usually fixed after training, they have less robustness between different wearable devices or subjects.

To figure out these challenges, this paper proposes an algorithm, named **F**all-detection **E**nsemble **D**ecision **T**ree (FEDT), to give superior robustness and high performance of sensitivity and specificity for fall detection. We evaluate our method on three open datasets and a practical dataset. What's more, to take advantage of cloud computing resources, we propose a mobile cloud collaboration system for fall detection. It is implemented based on FEDT with a wearable device and cloud server. The system is composed of three stages: 1) the mobile stage: a light-weighted threshold method is used to roughly filter the suspicious fall data, as computing ability of wearable device is limited, 2) the collaboration stage: the tri-axial accelerometer data of suspected fall is transmitted to the cloud server and features are extracted at the same time, 3) the cloud stage: the model trained by FEDT algorithm gives the final detection result with extracted features in 2).

The rest of the paper is organized as follows. In the section II, we review the related work of fall detection. In the section III, we present the proposed FEDT algorithm and the system in detail. In the section IV, we evaluate the performance of the FEDT. In the section V, the conclusion and future work are presented.

## II. RELATED WORK

In recent years, many researchers have paid close attention to fall detection. The main methods of fall detection can be divided into three categories: vision-based, ambient sensor based, wearable sensor based. In this section, we will review the methods of these work.

## A. Vision-Based Methods

Vision-based fall detection methods are based on videos and images from cameras. In [14], they use a MapCam (omni-camera) to capture images of a fixed area and extract the height and width of the subject in the images. Comparing the two characteristic with personal information of the subject, they can detect falls with a threshold method. [15] uses multiple cameras to detect falls. They reconstruct the 3-D shape of the subject. Then they analyze the volume through the vertical axis, and once most of the volume of the body is near the ground for a while, they judge the subject has fallen and alert for help. In [16], they use a single camera to capture images and detect falls by analyzing the velocity change of motion magnitude, human shape and motion orientation through processing the images. However, on the one hand, the cameras are usually set in a fixed place, which is not flexible, on the other hand, the problem of privacy is also an open issue to be settled. Additionally, storing and processing videos and images are costly.

## B. Ambient Sensor based Methods

The ambient sensors take advantages of the information of surroundings, such as Wi-Fi, ultrasonic, radar, infrared ray. In [5], they implement a fall detection system with a commodity Wi-Fi devices. They use the phase and amplitude from the Channel State Information (CSI) to detect falls. [6] places ultrasonic sensors on top and wall of room. It detects falls by comparing the signal between side and top signals with the threshold signals predefined. [7] uses the time-frequency features from radar data, and feed them into a deep learning model which is composed of stacked auto-encoders and a softmax layer as output. Then they use the model detect falls. [8] uses infrared array sensor to detect temperature distribution and extracts four features from the sensor readings. Then it gives fall detection result with a classifier (e.g. support vector machine, k-nearest neighbors). However, the ambient sensor is not target-specific, as it collects all the information of surroundings. The privacy issue is also concerned in these methods.

## C. Wearable Sensor based Methods

Wearable sensor based methods usually utilize the inertial sensors as accelerometer and gyroscope. [17] places several sensors on the subjects' chest and their feet to collect data and they propose a classifier which combines domain knowledge and machine learning method to detect falls. [18] puts an accelerometer on user's waist to detect different types of activities of people, such as run, walk, fall, etc. They use K-Nearest Neighbors (KNN) algorithm, and neural network algorithm to classify activities of people. In [19], they propose a fall detection alerting system with the tri-axial accelerator and gyroscope sensor in a mobile smart phone. Once the synthetic acceleration and total deflection angle are larger than thresholds defined before, the system will alert emergency center or the family members through messages with Global Position System (GPS) location or a call. However, these methods can't provide high sensitivity and specificity. [11] proposes a two-stage weight Extreme Learning Machine (ELM) method, which can achieve higher sensitivity and specificity but the robustness of this method is poor. There are many other machine learning methods which have been adopted in wearable sensor based methods, such as one class SVM (OCSVM) [20], HMM [21], et al. Although the wearable sensor based methods have developed a lot in recent years, there are much room for improvements in sensitivity, specificity and robustness.

Wearable sensor based methods are more common for great flexibility and ease of implementation. What's more, they are more target-specific and provide more private security, while the other two may retain other subjects' information.

## III. PROPOSED METHOD AND SYSTEM

In this section, we will explain FEDT algorithm and mobile cloud collaboration system explicitly.

The implemented system is composed of three stages: 1) in the mobile stage, light-weighted threshold method is adopted to roughly filter out the ADLs. If the threshold method judges the subjective to be suspected fall, it will go to the collaboration stage, or it will stay in the mobile stage. 2) in the collaboration stage, it will transmit acceleration data to cloud server, and the cloud server will extract features from received data. After collaboration stage, the system will directly go to the cloud stage. 3) in the cloud stage, the system will give final detection result through the model trained by FEDT with the extracted features in 2). Once the cloud detects the falls, it will return a warning to mobile device, which will alert for help. Fig. 1 shows the workflow of the system.

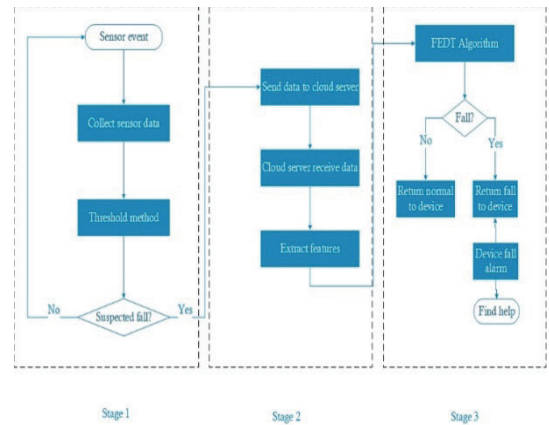

Fig. 1. The workflow of the system

## A. Mobile Stage: Threshold Method

In the mobile stage, tri-axial accelerometer sensor data is collected as x, y, z. The threshold method is used to roughly filter out the suspicious falls. In the training set, the root mean square (RMS) of tri-axial accelerometer sensor data is employed to filter out the ADLs, shown in Eq. 1. The threshold is set by the statistics of the training set, which should retain all the fall instances in the training dataset, and meanwhile filter out the ADLs as much as possible. The threshold method is chosen in the mobile stage, for it is a relatively light-weighted but a useful way to filter out the ADLs with less computing resources and storage, and it can be easily implemented in a mobile device.

$$\text{RMS} = \left(\sqrt{x^2 + y^2 + z^2}\right) \tag{1}$$

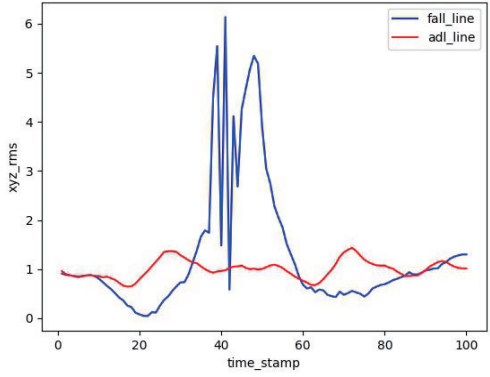

Fig. 2.  RMS of fall data and ADL data

Fig. 2 shows the RMS of ADL data and fall data. As we can see that peak value of fall data's RMS is much larger than the ADL's. To exploit this distinction, we design a threshold method, given in Fig. 3. If the RMS of collected data is larger than the threshold predefined, the system will turn to the collaboration stage, or the system will continue to run in the mobile stage.

**Algorithm 1 ThresHold**
**Input:** threshold, x, y, z
1: **function** THRESHOLD(threshold, x, y, z)
2:     $rms \leftarrow \sqrt{x^2 + y^2 + z^2}$
3:     **if** rms < threshold **then**
4:         continue to run in the mobile stage;
5:     **else**
6:         turn to the collaboration stage;
7:     **end if**
8: **end function**

Fig. 3.  Threshold method

### B. Collaboration Stage: Mobile Cloud Collaboration and Feature Extraction

TABLE I.
REPRESENTATIVE FEATURES AND DESCRIPTION

| Feature | Calculate Formula and Description |
|---|---|
| fft_coefficient | $C_k = \sum_{m=0}^{n-1} a_m exp\left\{-2\pi i \frac{mk}{n}\right\}$, $k = 0, \dots, n-1$ |
| abs_energy | $E = \sum_{i=1,\dots,n} t_i^2$ |
| absolute_changes | $Changes = \sum_{i=1,\dots,n-1} |t_{i+1} - t_i|$ |
| energy_ration _by_chunks | Calculate the sum of squares of chunk i out N chunks expressed as a ratio with the sum of squares over the whole series |
| first_location _of_maximum | Calculate the first location of the maximum of the time series data |

In the collaboration stage, the system will transmit tri-axial acceleration data to cloud server, and it will extract features from the received data in the cloud server. Here we use **FeatuRe Extraction** based on **Scalable Hypothesis** tests (FRESH algorithm) [22] to extract features from time series data. This algorithm can extract 794 features in total. The most representative features for fall detection are shown in TABLE I. The FRESH algorithm can be implemented in parallel, so it can extract features efficiently, which is important for a real-time fall detection system.

In this stage, Transmission Control Protocol (TCP) is adopted to transfer our data to the cloud. We choose the TCP rather than the User Datagram Protocol (UDP), because the UDP is not reliable, and may lose data during the transmission which may contain crucial information.

### C. Cloud Stage: Fall-Detection Ensemble Decision Tree (FEDT)

In the cloud stage, the system uses a model trained by FEDT algorithm to give final detection result. The cloud stage is implemented in the cloud server side, which has more computing resources and ability to give higher sensitivity and specificity detection result.

A tree ensemble model usually contains plenty of weak learners which are called as base learners in an ensemble model. The ensemble model's prediction is the sum of all base learners' prediction, as shown in Eq. (2).

$$y'_i = \varphi(x_i) = \sum_{m}^{M} f_m(x_i), f_m \in F \quad (2)$$

The $F$ is a collection of the base learners, $F = \{f(x) = S_{w(x)}\} (w : R^n \to K, S \in R^K)$, and each base learner in a tree ensemble model is a decision tree, which is usually called as **C**lassification **A**nd **R**egression **T**ree (CART). K represents how many leaf nodes in a tree while $S_{w(x)}$ means the score of x in a tree.

We define the objective function of the FEDT model as the following, given in Eq. (3). In the FEDT model, we will minimize the objective function.

$$Obj(\theta) = \sum_j l_1(y'_j, y_j) + \sum_m l_2(f_m) \quad (3)$$

Here the $l_1$ is a loss function to measure the difference between the ground truth and the prediction. In the Eq. (3), $y_j$ means the ground truth while the $y'_j$ means the prediction of the model. The $l_2$ is another loss function, which is used to control the complexity of the FEDT model. Once the model becomes complex, it may meet the problem of overfitting which is a tough problem in machine learning field. The $l_2$ can be defined in many ways. Here we define the $l_2$ in the following Eq. (4).

$$l_2 = \alpha K + \beta \sum_{k}^{K} w_k^2 \quad (4)$$

The α and β are hyper parameters in the FEDT model which are used to tune the model for better performance.

### D. System Implementation

We implement the system with Samsung watch gear s3 and a cloud server. The mobile stage is implemented in the watch. If the watch detects suspected fall, it will go to the collaboration stage, and then the cloud stage gives final detection result. The application demo is shown in Fig. 4. When the application receives fall detection signal from cloud server, it will alert for help. The third line shows the detection of the mobile stage, while the fourth line shows the detection of the cloud stage.

With this application, we also collected 252 fall samples, and more than 1000 ADL samples and the data forms our own practical dataset.

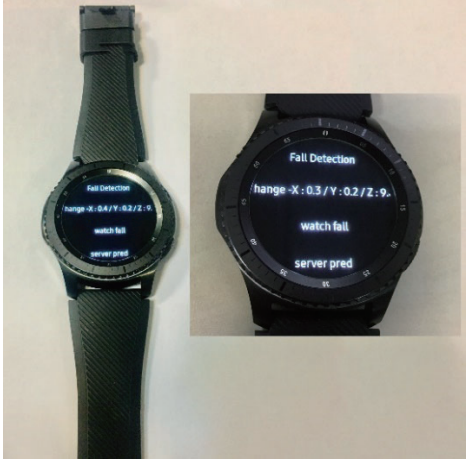

Fig. 4. Application demo on Samsung watch gear s3

## IV. Experiments

In this section, we will deploy sufficient experiments to show the effectiveness of the proposed FEDT.

### A. Datasets

In order to evaluate the performance of the proposed FEDT, three open datasets: SisFall [23], MobiAct [24], MMsys [25], and a practical dataset are used.

SisFall [23]: The SisFall dataset recorded 38 subjects' activities which were composed of elderly people and young adults. It contained 15 different types of falls and 19 different ADLs. It was recorded by a self-developed embedded device with the accelerometer and gyroscope data.

MobiAct [24]: The MobiAct dataset recorded 57 subjects' activities in the predefined activities. It contained 4 types of falls and 9 different ADLs with more than 2500 trials. It was recorded by a smart phone with the accelerometer, gyroscope, and orientation data.

MMsys [25]: The MMsys dataset recorded 42 subjects' activities. It contained 4 different types of falls and 11 different ADLs. It was recorded by two sensor nodes strapped to the chest and thigh of subjects with the accelerometer and gyroscope data.

Practical dataset: The practical dataset recorded 36 subjects' activities. It contained 5 different types of falls and 21 types of ADLs. It was recorded by Samsung watch gear s3 with accelerometer data.

### B. Preprocessing

To get the labeled data, we need to preprocess and segment the datasets. The segmentation strategy is different between the falls' data and the ADLs'. To get the labeled fall data, we first get the index of peak value in the RMS of the tri-axial accelerometer data. And then we choose some data from the left and right of the index. The length of the left is usually equal to the right's. To get the labeled ADL data, we adopt sliding window mechanism to segment the data. Fig. 5 shows the segmentation strategy of fall and ADL data.

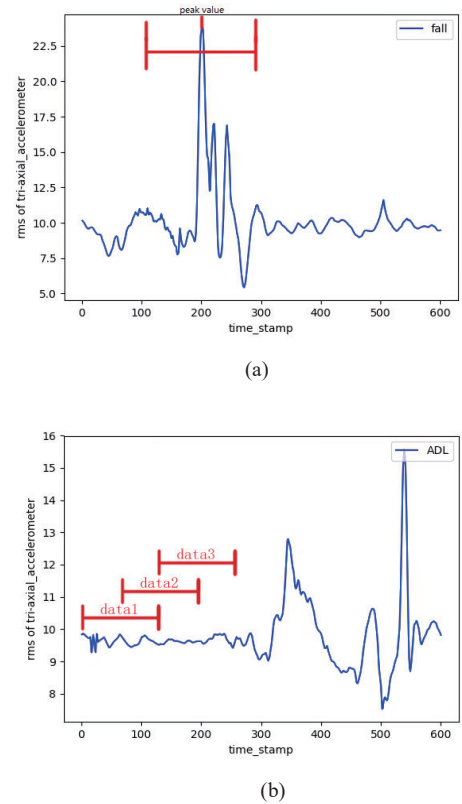

Fig. 5. Figure (a) is the seamentation stragegy of fall data. Figure (b) is the sliding window mechanism of ADL data.

When segmenting the ADL data, we use different window size in different dataset. In the SisFall, the size is 200 while in the MMsys it is set as 100. The size of MobiAct is 600, and the size of our practical dataset is 300. The amount of the fall and ADL samples after segmentation is shown in TABLE II.

TABLE II.
THE AMOUNT OF FALL AND ADL SAMPLES AFTER SEGMENTATION

| Dataset | Fall | ADL | Amount |
|---|---|---|---|
| SisFall | 1798 | 52066 | 53864 |
| MMsys | 416 | 43866 | 44282 |
| MobiAct | 767 | 50857 | 51624 |
| Practical | 252 | 27452 | 27704 |

### C. Experimental Analysis

We evaluate the FEDT algorithm on the four datasets. Here 10-fold cross validation is used to evaluate the FEDT algorithm. The evaluation metrics are defined in Eq. (5), (6), (7), (8).

$$sensitivity = \frac{TP}{TP + FN} \quad (5)$$

$$specificity = \frac{TN}{TN + FP} \quad (6)$$

$$precision = \frac{TP}{TP+FP} \quad (7)$$

$$f1\text{-}score = \frac{2 * precision * sensitivity}{precision + sensitivity} \quad (8)$$

The performance of the FEDT algorithm is reported in TABLE III. From the table we can see that the FEDT algorithm can give high sensitivity and specificity which is important in fall detection. High specificity means low false positive rate of fall detection in daily living activities

TABLE III.  RESULTS OF FEDT ALGORITHM ON FOUR DATASETS

| Datasets  | Sensitivity | Specificity | Precision | F1-score |
|-----------|-------------|-------------|-----------|----------|
| SisFall   | 98.11%      | 99.98%      | 99.66%    | 98.87%   |
| MMsys     | 97.33%      | 99.97%      | 97.38%    | 97.27%   |
| MobiAct   | 98.05%      | 99.95%      | 99.74%    | 98.87%   |
| Practical | 95.25%      | 99.98%      | 98.46%    | 96.78%   |

We compared the FEDT algorithm with random forest and GBDT as the three algorithms are all ensemble learning methods. The performance of the three algorithms is shown in Fig. 6. From the results, it is obviously that the proposed FEDT outperforms the other two algorithms both on sensitivity and specificity. Especially in the practical dataset, our FEDT outperforms significantly. We analyze that the FEDT has the $l_2$ loss function to avoid the overfitting problem while the GBDT and random forest are easy to overfit the training data without the $l_2$ loss function. Besides, the random forest ignores the relevance between the features, thus making it may loss important information. What's more, the random forest resembles a black box, which we don't know how it works internally, and we just can tune the model by random seed and parameters.

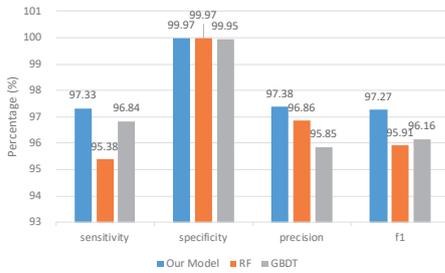

(a)

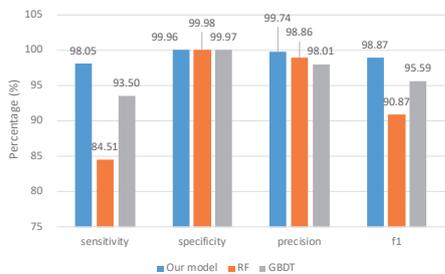

(b)

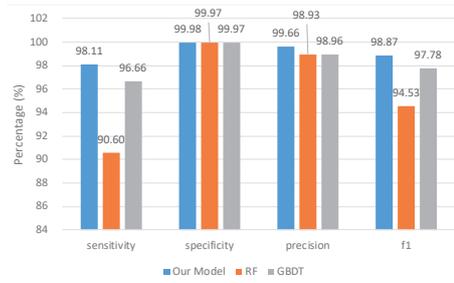

(c)

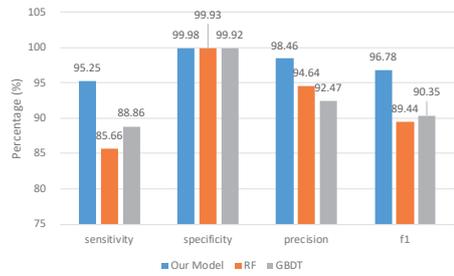

(d)

Fig. 6. Figure (a) is the results on MMsys dataset. Figure (b) is the results on MobiAct dataset. Figure (c) is the results on SisFall dataset. Figure (d) is results on practical dataset

Furthermore, we compare our method with some of the state-of-the-art methods on different datasets. To avoid the bias when rebuilding these methods, we evaluate these methods only on the datasets adopted in their papers. On MobiAct dataset, we choose [26], [27], [28], on MMsys dataset, we select [29], [30], [31] and on SisFall dataset, we choose [32], [33], [34]. Results are shown in TABLE IV. TABLE V. TABLE VI.

TABLE IV.  SENSITIVITY AND SPECIFICITY ON MOBIACT

|             | Chatzaki et al [26] | Jahanjoo et al [27] | Tsinganos et al [28] | FEDT model |
|-------------|---------------------|---------------------|----------------------|------------|
| Sensitivity | 75.10%              | 97.29%              | 95.52%               | **98.05%** |
| Specificity | 89.79%              | 98.70%              | 97.07%               | **99.95%** |

TABLE V.  SENSITIVITY AND SPECIFICITY ON MMSYS

|             | Meng et al [29] | Khan et al [30] | Putra et al [31] | FEDT model |
|-------------|-----------------|-----------------|------------------|------------|
| Sensitivity | 91.08%          | 90.50%          | 93.60%           | **97.33%** |
| Specificity | 87.45%          | 60.60%          | 92.30%           | **99.97%** |

TABLE VI.  SENSITIVITY AND SPECIFICITY ON SISFALL

|             | Sucerquia et al [32] | Carletti et al[33] | Nguyen et al[34] | FEDT model |
|-------------|----------------------|--------------------|-------------------|------------|
| Sensitivity | 97.35%               | 91.20%             | **99.73%**        | 98.11%     |
| Specificity | 91.49%               | 98.10%             | 97.70%            | **99.98%** |

These three tables show that the FEDT gives the best performance of sensitivity and specificity on both MobiAct and MMsys dataset. While the FEDT gives the best specificity while the sensitivity is close to the best on SisFall dataset. Compared with these conventional methods, the FEDT is an ensemble learning method which can avoid underfitting problem and have better learning capacity.

*D. Dimension Reduction Experiments*

Although the 794 features can be extracted efficiently through the FRESH algorithm, we also try to adopt dimension reduction techniques to reduce the dimension of the features. Here we use the Principal Components Analysis (PCA) to reduce the dimension of the features. We evaluate the PCA on the SisFall dataset and MMsys dataset. We keep the 95% of the principal components. The results are shown in TABLE VII. Both in the SisFall dataset and MMsys dataset, we can see that sensitivity drops a lot with the PCA technique.

As the PCA breaks the crucial features for classification, the new features are difficult to distinguish after using the PCA, thus making it hard to classify the fall and ADLs. So, according to the results, we can't reduce the feature dimension through PCA in order to keep the performance of the FEDT algorithm.

TABLE VII.
SENSITIVITY AND SPECIFICITY AFTER PCA TECHNIQUE

| Datasets | Sensitivity | Specificity | Precision | F1-score |
|---|---|---|---|---|
| SisFall (PCA) | 73.30% | 99.84% | 94.15% | 82.31% |
| SisFall | **98.11%** | 99.98% | 99.66% | 98.87% |
| MMsys (PCA) | 65.61% | 99.95% | 92.77% | 76.49% |
| MMsys | **97.33%** | 99.97% | 97.38% | 97.27% |

*E. Robustness Experiments*

As the sensor differs from device to device and the subjects are different, the data collected by different devices and subjects have different distributions. To evaluate the robustness of the FEDT, we collect a new practical dataset by Huawei smartphone. We use the Huawei dataset to measure the performance of the model trained with the dataset collected by Samsung watch. In this experiment, there are 97 samples in Huawei dataset which are composed of 55 falls and 42 ADLs. The results are shown in the Fig. 7.

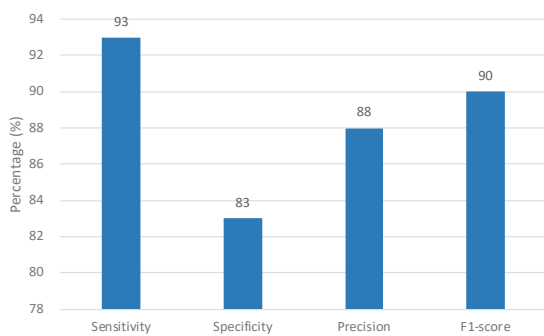

Fig. 7. Results of robustness experiments

The results report that the FEDT has superior robustness in sensitivity. As the FEDT is an ensemble algorithm, the detection result is determined by not only a classifier but the whole classifiers. Through the ensemble method, the model trained by FEDT algorithm has more transfer ability between different datasets.

V. CONCLUSION

In this paper, we propose a novel fall detection algorithm named FEDT based on decision tree and a mobile cloud collaboration system. The experimental results show that the proposed FEDT algorithm gives better performance than most of the fall detection algorithms and has superior robustness between different datasets.

Even though we can achieve 93% fall detection sensitivity in the robustness experiments, in the future, we will upgrade this algorithm to device-free manner with transferring learning technique, so as to further improve the detection performance. As we can see from the TABLE VII. , the PCA is not serviceable for reducing the dimension of the features. In the feature, we will try other feature extraction algorithms and dimension reduction techniques to reduce the dimension of the features and keep the performance of the FEDT meanwhile.

VI. ACKNOWLEDGMENTS

This research is Supported by Beijing Natural Science Foundation (4194091).